\documentclass[singlecolumn,superscriptaddress,preprintnumbers,amsmath,prb,amssymb]{revtex4}

\usepackage{graphicx}
\usepackage{dcolumn}
\usepackage{bm}

\newcommand{\bra}[1]{\langle #1|}
\newcommand{\ket}[1]{|#1\rangle}

\newcommand{\beq}{\begin{eqnarray}}
\newcommand{\eeq}{\end{eqnarray}}

\begin{document}

\title{Electrical current and coupled electron-nuclear
spin dynamics in double quantum dots}

\author{G. Giavaras}
\affiliation{Advanced Science Institute, RIKEN, Wako-shi, Saitama
351-0198, Japan}

\author{Neill Lambert}
\affiliation{Advanced Science Institute, RIKEN, Wako-shi, Saitama
351-0198, Japan}

\author{Franco Nori}
\affiliation{Advanced Science Institute, RIKEN, Wako-shi, Saitama
351-0198, Japan} \affiliation{Department of Physics, The
University of Michigan, Ann Arbor, MI 48109-1040, USA }


\begin{abstract}
We examine electronic transport in a spin-blockaded double quantum
dot. We show that by tuning the strength of the spin-orbit
interaction the current flowing through the double dot exhibits a
dip at zero magnetic field or a peak at a magnetic field for which
the two-electron energy levels anticross. This behaviour is due to
the dependence of the singlet-triplet mixing on the field and
spin-orbit amplitude. We derive approximate expressions for the
current as a function of the amplitudes of the states involved in
the transport. We also consider an alternative model that takes
into account a finite number of nuclear spins and study the
resulting coupled dynamics between electron and nuclear spins. We
show that if the spin ensemble is in a thermal state there are
regular oscillations in the transient current followed by
quasi-chaotic revivals akin to those seen in a thermal
Jaynes-Cummings model.
\end{abstract}

\maketitle

\section{Introduction}

A double quantum dot (DQD) can be used for the detailed
investigation of spin interactions among electron spins and even
between electron and nuclear spins. The interactions can be probed
optically or electrically by monitoring the electrical current
flowing through the DQD as a function of the energy offset between
the two dots and the applied magnetic field. In a spin-blockaded
DQD, the current through the DQD is large when the two electrons
form a singlet state, whereas it is suppressed when the two
electrons form a triplet state.~\cite{ono} The spin blockade
mechanism is due to the Pauli principle and has been demonstrated
in semiconductor heterostructure quantum dots~\cite{ono} as well
as carbon nanotube dots.~\cite{buitelaar} It has also been shown
that the transient behaviour of the leakage current can provide
valuable information about the interactions between electrons in
the DQD and nuclear spins in the host material.~\cite{ono2004} In
particular, the electrical transport process can lead to a coupled
electron-nuclear dynamics, nuclear spin polarization, and
hysteresis effects.~\cite{ono,ono2004,abolfath}

In the spin blockade regime, the small leakage current increases
when there is a process that leads to singlet-triplet
hybridization (mixing). A non spin-conserving interdot tunnelling
is one such process. This type of tunnelling may result from a
spin-orbit interaction (SOI), which in some cases can be strong
enough and thus has to be taken into account.~\cite{pfund, perge}
The hyperfine interaction (HI) between dot electrons and nuclear
spins can also lead to singlet-triplet hybridization. In a
simplified approach, the nuclear spins create an effective
magnetic field which acts on the electron spins. This field can
point in an arbitrary direction and mixes singlet and triplet
states.

The electrical current in a DQD system has been examined
theoretically in the presence of strong SOI and in the regime
where the coupling of the DQD to the leads corresponds to the
largest rate in the system.~\cite{danon2009} In the first part of
this work, we consider a DQD weakly coupled to the leads and
examine the current in the resonant regime, i.e., when the lowest
singlet and triplet energy levels are almost aligned. Unlike the
approach followed in Ref.~\onlinecite{danon2009} we consider
explicitly the lowest one-electron states and derive rate
equations that involve the transition rates between one- and
two-electron states. The SOI is modelled with a non
spin-conserving tunnel coupling amplitude $t_{\mathrm{so}}$
between the two dots~\cite{danon2009, romeo} which couples the
triplet states $\ket{T_+}$, $\ket{T_-}$ with singlet states. We
consider weak SOI, that is $t_{\mathrm{so}}< t_{\mathrm{c}}$,
where $t_{\mathrm{c}}$ is the spin-conserving interdot tunnel
coupling. The effect of a strong SOI on a spin-blockaded DQD has
been investigated in Ref.~\onlinecite{danon2009} with a more
general SOI model in which all triplet states couple to singlet
states. Further, we here give emphasis to the regime where the HI
is weak enough or absent, which might be the case in carbon-based
quantum dots.~\cite{rozhkov, chorley} The main aim of the first
part of this work is to determine the current as a function of the
amplitudes of the one- and two-electron states which participate
in the transport cycle. Some approximate results which give
valuable insight into the basic behavior of the current are
derived. We show that, depending on the strength of the SOI, the
current shows a dip at zero magnetic field or a peak when the
lowest two-electron energy levels anticross. This behaviour occurs
because when $t_{\mathrm{so}}$ is large the singlet-triplet mixing
near zero field is much weaker compared to that at high field.
This gives rise to a dip at zero field. However, when
$t_{\mathrm{so}}$ is small the mixing is strong only near the
anticrossing point leading to a peak in the current.

In the second part of this work we focus mostly on the interplay
between SOI and HI. To properly account for the HI, we employ a
microscopic `toy' model which takes into account a finite number
of nuclear spins. To make this model tractable, only the two
lowest singlet-triplet states are considered, and we treat the
nuclear spins as a single large spin. We first consider the
interplay between the SOI and HI, and its effect on the steady
state transport and nuclear spin polarization. We find a sharp
transition in the current and polarization as the SOI is
increased, consistent with the topological phase transition
investigated in Ref.~\onlinecite{Rudner10}. Second, we look at the
transient dynamics induced by the HI alone, and find a strong
oscillatory contribution depending on the hyperfine coupling
strength and inversely proportional to the square root of the
number of nuclear spins.

\section{Electrical current in the spin blockade regime}

\subsection{Physical Model}

In this section the electrical current through the DQD is examined
when the electron-nuclear spin dynamics are uncoupled. This
section is concerned with the effect of the SOI on the electrical
current while HI-induced effects due to the coupled
electron-nuclear dynamics are addressed in the next section. The
DQD is modelled with the two-site Hamiltonian
\begin{equation}
H_{\mathrm{DQD}}=H_{\mathrm{c}}+H_{\mathrm{so}}+H_{\mathrm{hf}}+\sum_{i=1}^{2}\varepsilon_{i}n_{i}
+U\sum_{i=1}^{2}n_{i\uparrow}n_{i\downarrow}+\frac{\Delta}{2}\sum_{i=1}^{2}\sigma_{i}^{\mathrm{z}}.
\label{molecule}
\end{equation}
Here $H_{\mathrm{c}}$ is the tunnel-coupling Hamiltonian that
conserves spin and has the form
\begin{equation}
H_{\mathrm{c}}=-t_{\mathrm{c}}(c^{\dagger}_{1\uparrow}c_{2\uparrow}
+c^{\dagger}_{1\downarrow}c_{2\downarrow})+\mathrm{H.c.},
\end{equation}
and the Hamiltonian part due to the SOI that allows spin-flip has
the form~\cite{romeo}
\begin{equation}
H_{\mathrm{so}}=-t_{\mathrm{so}}(
c^{\dagger}_{1\uparrow}c_{2\downarrow} -
c^{\dagger}_{1\downarrow}c_{2\uparrow})+\mathrm{H.c.}.
\end{equation}
For the HI we assume the form~\cite{jouravlev}
\begin{equation}
H_{\mathrm{hf}}=\frac{1}{2} g_{\mathrm{e}} \mu_{\mathrm{B}}  (
{\bf B}_{\mathrm{N},1} \bm{\sigma}_{1}  + {\bf B}_{\mathrm{N},2}
\bm{\sigma}_{2} ).
\end{equation}
The operator $c_{i\sigma}^{\dagger}$ ($c_{i\sigma}$) creates
(destroys) an electron on dot $i=1,2$ with spin
$\sigma=\{\uparrow,\downarrow\}$ and orbital energy
$\varepsilon_{i}$. The number operator is denoted by
$n_{i}=\sum_{\sigma}n_{i\sigma}=c_{i\uparrow}^{\dagger}c_{i\uparrow}+c_{i\downarrow}^{\dagger}c_{i\downarrow}$.
The tunnel coupling amplitude between the two dots is
$t_{\mathrm{c}}$, the amplitude due to the SOI is
$t_{\mathrm{so}}$, the charging energy is $U$, and
$\Delta=g_{\mathrm{e}} \mu_{\mathrm{B}}B$ is the Zeeman splitting
due to the external magnetic field $B$ in the $z$-direction. Here
${\bf B}_{\mathrm{N},i}$ is the magnetic field in the $i$th dot
due to the nuclear spins and $\bm{\sigma}_{i}$ are the Pauli
operators.

The quantum states which participate in the transport cycle in the
spin blockade regime are the 2 lowest one-electron states and the
5 lowest two-electron states. For simplicity the three-electron
states are neglected in this study because they do not change
qualitatively the basic results. The one-electron eigenstates
$\ket{j;n=1}$ can be written in the general form
\begin{equation}
\ket{j;1}=\alpha_{j} \ket{\uparrow,0} + \beta_{j}\ket{0,\uparrow}
+ \gamma_{j}\ket{\downarrow,0} + \delta_{j}\ket{0,\downarrow},
\end{equation}
with $j=1,2$ and $c^{\dagger}_{1\sigma}\ket{0}=\ket{\sigma,0}$,
$c^{\dagger}_{2\sigma}\ket{0}=\ket{0,\sigma}$. Here the
eigenstates are ordered with increasing energy. In the
spin-blockade regime $U\gg t_{\mathrm{c}}$ and further there is an
energy offset between the two dots. In this work we choose for the
on site energies $\varepsilon_{2}=\varepsilon_{1}-U/2$ and define
the energy detuning as $\delta=E(1,1)-E(0,2)$, where $E(n,m)$ is
the energy of the charge state which has $n$ $(m)$ electrons on
dot $i=1$ ($i=2$). If $H_{\mathrm{so}}=0$ and $H_{\mathrm{hf}}=0$
then $\alpha_{1}=\beta_{1}=0$ and $\gamma_{2}=\delta_{2}=0$ and
the nonzero amplitudes satisfy $\delta_{1}\gg\gamma_{1}$ and
$\beta_{2}\gg\alpha_{2}$. When $H_{\mathrm{so}}\ne0$ the
amplitudes $\alpha_{1}$, $\beta_{1}$, $\gamma_{2}$, $\delta_{2}$
are in general nonzero and satisfy $\beta_{1}\ll\alpha_{1}$,
$\delta_{2}\ll\gamma_{2}$.

Neglecting double occupation on dot 1, a two-electron eigenstate
$\ket{j;n=2}$ with $j=3,...7$ has the general form
\begin{equation}
\ket{j;2}= a_{j} \ket{\uparrow,\uparrow} +
b_{j}\ket{\uparrow,\downarrow} + c_{j}\ket{\downarrow,\uparrow} +
d_{j}\ket{\downarrow,\downarrow} +
e_{j}\ket{0,\uparrow\downarrow}.\label{2states}
\end{equation}
Here
$\ket{T_{-}}=\ket{\downarrow,\downarrow}=c^{\dagger}_{1\downarrow}c^{\dagger}_{2\downarrow}\ket{0}$,
$\ket{T_{+}}=\ket{\uparrow,\uparrow}=c^{\dagger}_{1\uparrow}c^{\dagger}_{2\uparrow}\ket{0}$,
$\ket{\sigma,\sigma'}=c^{\dagger}_{1\sigma}c^{\dagger}_{2\sigma'}\ket{0}$,
and
$\ket{S_{02}}=\ket{0,\uparrow\downarrow}=c^{\dagger}_{2\uparrow}c^{\dagger}_{2\downarrow}\ket{0}$.
The amplitudes of the various components depend on the strengths
of the SOI and HI as well as the Zeeman splitting and detuning.
When $H_{\mathrm{so}}=0$ and $H_{\mathrm{hf}}=0$ the two-electron
eigenstates correspond to the triplet states $\ket{T_{+}}$,
$\ket{T_{-}}$,
$\ket{T_{0}}=(\ket{\uparrow\downarrow}+\ket{\downarrow\uparrow})/\sqrt{2}$
and the two singlet states, which consist of the components
$\ket{S_{02}}=\ket{0,\uparrow\downarrow}$ and
$\ket{S_{11}}=(\ket{\uparrow\downarrow}-\ket{\downarrow\uparrow})/\sqrt{2}$.
The effect of the $H_{\mathrm{so}}$ is to couple the
$\ket{T_{+}}$, $\ket{T_{-}}$ states to singlet components. As
shown below the coupling strength increases with $t_{\mathrm{so}}$
and, for a fixed detuning, is sensitive to the Zeeman splitting.

To calculate the electrical current flowing through the DQD we
employ a density matrix approach within the Born and Markov
approximations.~\cite{gardiner} The internal parameters of the DQD
and the chemical potentials of the two leads are adjusted to the
spin-blockade regime. This regime can be identified from the fact
that when $H_{\mathrm{so}}=0$ and $H_{\mathrm{hf}}=0$ the three
triplet states $\ket{T_{+}}$, $\ket{T_{-}}$, and $\ket{T_{0}}$ are
equally and almost fully populated ($\sim1/3$), provided spin
decoherence is ignored, and the current as a function of the
source-drain bias is suppressed.

\subsection{Results}

\begin{figure}
\begin{center}
\includegraphics[height=8.50cm,angle=0]{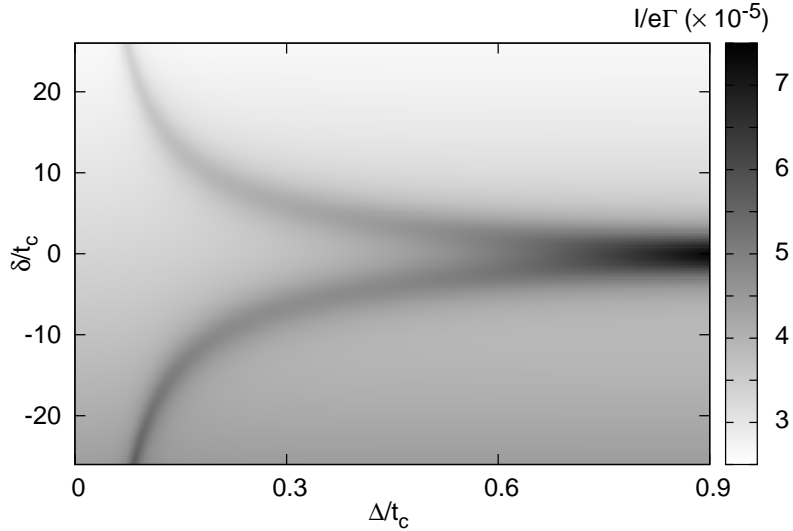}\\
\caption{Electrical current, $I$, as a function of the Zeeman
splitting, $\Delta$, and energy detuning, $\delta$, when there is
no hyperfine interaction. The SOI amplitude is
$t_{\mathrm{so}}=0.01t_{\mathrm{c}}$, where $t_{\mathrm{c}}$ is
the spin-conserving interdot tunnel coupling.}\label{current2d}
\end{center}
\end{figure}

When $t_{\mathrm{so}}$ is nonzero the $\ket{T_{+}}$, $\ket{T_{-}}$
states couple to singlet states and this coupling has a direct
effect on the current. To demonstrate this effect we show in
Fig.~\ref{current2d} the current as a function of the Zeeman
splitting $\Delta$ and energy detuning $\delta$ between the two
dots when there is no HI. The coupling of the states is strong
near the anti-crossing points leading to an increase in the
current (see also below). As a result the curves of high current
map-out the points where the energy levels of the quasi singlet
and triplet states anticross. When $t_{\mathrm{so}}=0$ the leakage
current is approximately constant and no high current curves
occur.

\begin{figure}
\begin{center}
\includegraphics[height=11.50cm,angle=270]{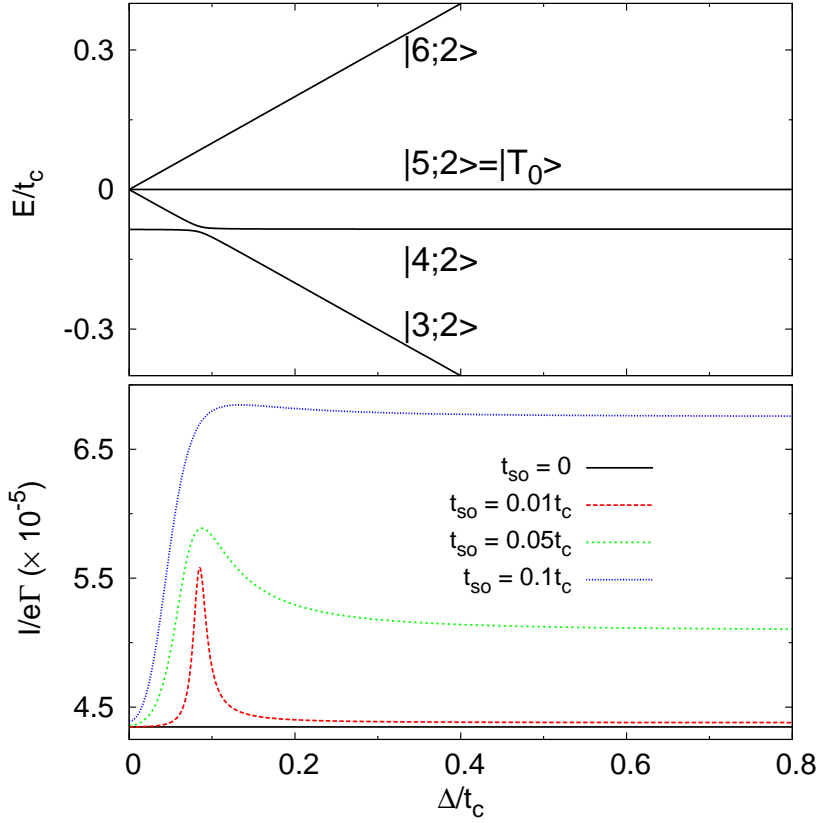}
\caption{ (Color online) The top panel shows the energies, $E$, of
the four lowest two-electron states as a function of the Zeeman
splitting, $\Delta$, for an energy detuning
$\delta=-25t_{\mathrm{c}}$ and SOI amplitude
$t_{\mathrm{so}}=0.1t_{\mathrm{c}}$, where $t_{\mathrm{c}}$ is the
spin-conserving interdot tunnel coupling. The notation of the
states is given in the main text in Eq.~(\ref{2states}). The
bottom panel shows the electrical current, $I$, as a function of
the Zeeman splitting $\Delta$ for an energy detuning
$\delta=-25t_{\mathrm{c}}$.}\label{current1d}
\end{center}
\end{figure}

To understand the role of the SOI we show in Fig.~\ref{current1d}
the energy diagram of the two-electron states as well as the
current as a function of the Zeeman splitting $\Delta$, for a
fixed energy detuning $\delta$ and different SOI amplitudes with
$H_{\mathrm{hf}}=0$. We concentrate on the regime
$t_{\mathrm{so}}<t_{\mathrm{c}}$ and choose $\delta<0$, which is
experimentally the most interesting case for spin-qubit
applications~\cite{ono, chorley, petta}, because the spin pair can
be described by an effective Heisenberg model. The resonant
current that is defined as the current at the anticrossing point
increases with $t_{\mathrm{so}}$ and the same occurs for the
asymptotic current that is defined as the current at a high
magnetic field. Therefore, when $t_{\mathrm{so}}$ is large enough
the asymptotic current becomes approximately equal to the resonant
current and thus the peak cannot be distinguished. The same
pattern occurs when the direction of the magnetic field is
reversed, thus the current as a function of the Zeeman splitting
shows a dip at $\Delta=0$. To a good approximation this pattern is
independent of the detuning, provided $t_{\mathrm{so}}$ is large,
and consequently the anticrossing points of the energy diagram
cannot be probed.

To quantify the above results we analyze the rate equations and
calculate analytically the transition rates. Then we derive the
steady-state current for some interesting limits, such as for
example the current at the singlet-triplet anticrossing point, as
well as for $B\sim 0$ and $B$ high. We are interested in
determining the current in the steady-state for a DQD that is
weakly coupled to the leads. In this regime we can consider only
the diagonal elements of the density matrix~\cite{newreference}
and the dynamics of the system is described by the rate equations
\begin{equation}
\frac{d\rho_{n}}{dt}=-\rho_{n}\sum_{m} R_{nm} +
\sum_{m}\rho_{m}R_{mn},\label{rateequations}
\end{equation}
where the diagonal elements of the density matrix are denoted by
$\rho_{n}=\rho_{nn}$ and the normalization condition is
$\sum^{7}_{n=1}\rho_{n}=1$. The transition rate from an eigenstate
$\ket{n}$ of the DQD with eigenenergy $E_{n}$ to an eigenstate
$\ket{m}$ with eigenenergy $E_{m}$ is
\begin{eqnarray}
R_{nm} &=& \Gamma\sum_{\sigma}[ |\langle
n|c_{1\sigma}|m\rangle|^{2}f_{L}(E_{mn}) +|\langle
m|c_{1\sigma}|n\rangle|^{2}f^{-}_{L}(E_{nm})\nonumber\\
&+&|\langle n|c_{2\sigma}|m\rangle|^{2}f_{R}(E_{mn}) +|\langle
m|c_{2\sigma}|n\rangle|^{2}f^{-}_{R}(E_{nm}) ] ,
\end{eqnarray}
where $\Gamma=2\pi|t_{r}|^{2}D_{r}/\hbar$. The tunnelling
amplitude between dot and lead is $t_{r}$, $f_{\ell}(E_{mn})$ is
the Fermi-Dirac distribution function at chemical potential
$\mu_{\ell}$, with $E_{mn}=E_{m}-E_{n}$ and $f^{-}=1-f$. Also,
$D_{r}$ is the density of states for the leads, which we assume to
be constant and equal for both leads.

\begin{figure}
\begin{center}
\includegraphics[height=11.50cm,angle=270]{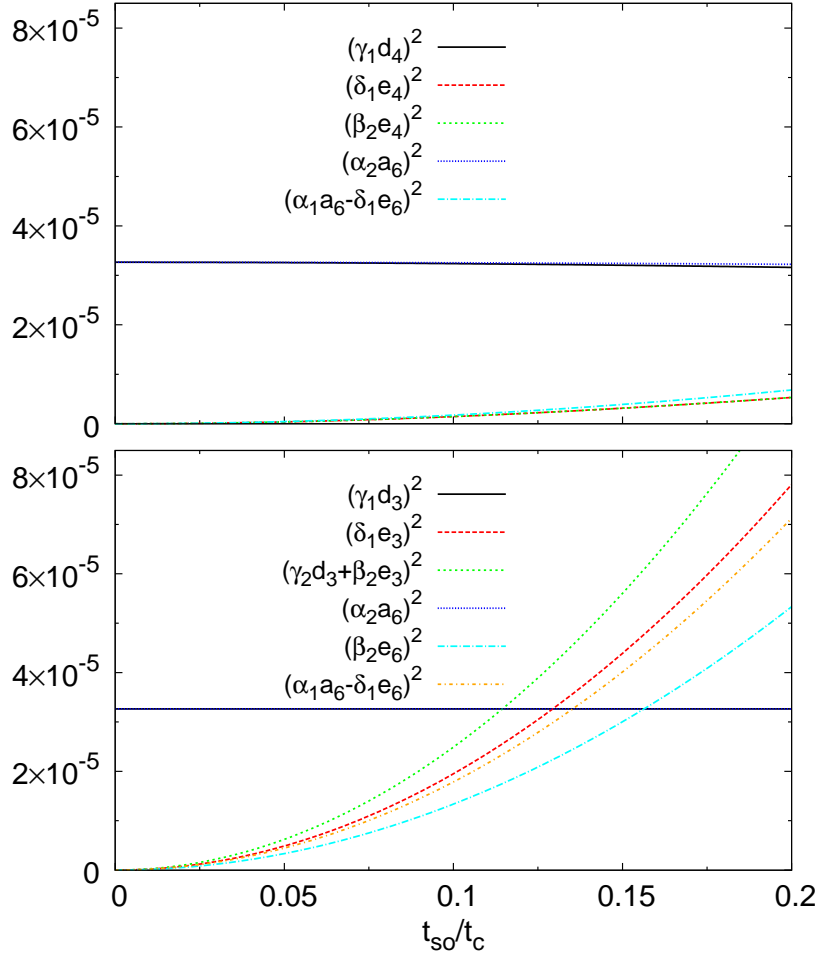}
\caption{ (Color online) The figure shows the important quantities
which determine the transition rates through the parameters
$F_{0}$, $E_{0}$ (top panel) and $F_{a}$, $E_{a}$ (bottom panel)
defined in Eqs.~(\ref{e0}), (\ref{f0}) and Eqs.~(\ref{ea}),
(\ref{fa}) respectively. The energy detuning is
$\delta=-25t_{\mathrm{c}}$ and the Zeeman splitting is chosen in
the low field regime $\Delta=0.02t_{\mathrm{c}}$ (top) and
asymptotic regime $\Delta=0.7t_{\mathrm{c}}$
(bottom).}\label{parameters}
\end{center}
\end{figure}

The operator for the electrical current, for example, for the
right lead is
\begin{equation}
\hat{I}= \frac{e}{\hbar}i\sum_{k\sigma}
t_{r}(c^{\dagger}_{2\sigma} d_{{k}\sigma} -
d^{\dagger}_{{k}\sigma} c_{2\sigma}),
\end{equation}
where $d_{{k}\sigma}$ denotes a lead operator. Tracing out the
leads we derive the following expression for the average current
\begin{equation}
I=e\Gamma\sum_{nm\sigma}\rho_{n} [ |\langle
n|c_{2\sigma}|m\rangle|^{2} f_{R}(E_{mn}) - |\langle
m|c_{2\sigma}|n\rangle|^{2} f^{-}_{R}(E_{nm})].
\end{equation}
Starting with the rate equations and calculating the transition
rates it can be readily derived that the absolute value of the
current for $H_{\mathrm{hf}}=0$ is $I = e \sum^{7}_{n=3}
(R_{n1}+R_{n2})\rho_{n}$. The simplest regime is when
$t_{\mathrm{so}}=0$ and $B=0$. Here, the occupations of the
triplet states satisfy $\rho_{4}=\rho_{5}=\rho_{6}=\rho_{T}$. From
the steady-state condition, $d\rho_{n}/dt=0$, it can be derived
that
\begin{equation}
3\rho_{T} \approx 1 - \left(
\frac{R_{14}}{R_{41}+R_{42}}+\frac{R_{15}}{R_{51}+R_{52}}+
\frac{R_{16}}{R_{61}+R_{62}}\right)^{-1}.
\end{equation}
For weakly coupled dots the second term is typically negligible
and $\rho_{T}\approx 1/3$. It is easy to prove that this
approximation is excellent when $t_{c}$ is small and $\delta$ is
large. In this regime the leakage current is
\begin{equation}
I_{T} \approx 2 e \Gamma \left(
\frac{R_{14}}{R_{41}+R_{42}}+\frac{R_{15}}{R_{51}+R_{52}}+
\frac{R_{16}}{R_{61}+R_{62}}\right)^{-1}.
\end{equation}
For $t_{\mathrm{so}}\neq 0$ we calculate analytically the
transition rates and derive that for low magnetic fields the
current is given approximately by the expression
\begin{equation}
I_0 \approx
\frac{2e\Gamma}{\frac{1}{2}F_{0}+\frac{1}{2}E_{0}},\label{I0}
\end{equation}
where
\begin{equation}
E_0=\frac{1}{ (\gamma_{1}d_{4} )^2 + (\delta_{1}e_{4})^2 +
(\beta_{2} e_{4})^2 }+ \frac{R_{15}}{R_{51}+R_{52}},\label{e0}
\end{equation}
\begin{equation}
F_0=\frac{1}{ (\alpha_{2}a_{6})^2 + ( \alpha_{1}a_{6} -
\delta_{1}e_{6} )^2 } + \frac{R_{25}}{R_{51}+R_{52}}.\label{f0}
\end{equation}
The transition rates $R_{15}$, $R_{25}$ involve the one-electron
state $\ket{1;1}$, $\ket{2;1}$ respectively, and the triplet state
$\ket{T_{0}}=\ket{5;2}$. The terms which are proportional to
$R_{12}$, $R_{25}$ do not affect the physics we examine here, so
for this reason they are not given explicitly. Also, as seen in
Fig.~\ref{current1d}, in the small $t_{\mathrm{so}}$ regime that
we are interested in $I_{0}$ is to a good approximation
independent of $t_{\mathrm{so}}$. An approximate expression for
the resonant current is
\begin{equation}
I_r \approx \frac{2e\Gamma}{\frac{2}{3}F_r+\frac{1}{3}E_{r}},
\end{equation}
with the parameters
\begin{equation}
E_r=\frac{1}{ (\gamma_{1}d_{4} )^2 + (\delta_{1}e_{4})^2 + (
\beta_{2} e_{4} )^2  }+ \frac{R_{15}}{R_{51}+R_{52}},\label{er}
\end{equation}
\begin{equation}
F_r=\frac{1}{ (\alpha_{2}a_{6})^2 + (\beta_{2}e_{6})^2 + (
\alpha_{1}a_{6} - \delta_{1}e_{6} )^2 } +
\frac{R_{25}}{R_{51}+R_{52}}.\label{fr}
\end{equation}
The resonant current corresponds to the magnetic field $B_{r}$ for
which the lowest quasi singlet and triplet states anticross and it
is well-defined for $t_{\mathrm{so}}\ll t_{\mathrm{c}}$. The
asymptotic current that corresponds to a high $B$ is given
approximately by the expression
\begin{equation}
I_a \approx
\frac{2e\Gamma}{\frac{1}{2}F_{a}+\frac{1}{2}E_{a}},\label{Ia}
\end{equation}
with
\begin{equation}
E_a=\frac{1}{ (\gamma_{1}d_{3} )^2 + (\delta_{1}e_{3})^2 + (
\gamma_{2}d_{3}+\beta_{2} e_{3} )^2 } +
\frac{R_{15}}{R_{51}+R_{52}},\label{ea}
\end{equation}
\begin{equation}
F_a=\frac{1}{ (\alpha_{2}a_{6})^2 + (\beta_{2}e_{6})^2 + (
\alpha_{1}a_{6} - \delta_{1}e_{6} )^2 } +
\frac{R_{25}}{R_{51}+R_{52}}.\label{fa}
\end{equation}
The asymptotic current is defined at a high magnetic field where
the current varies slowly with $B$.

\begin{figure}
\begin{center}
\includegraphics[height=11.50cm,angle=270]{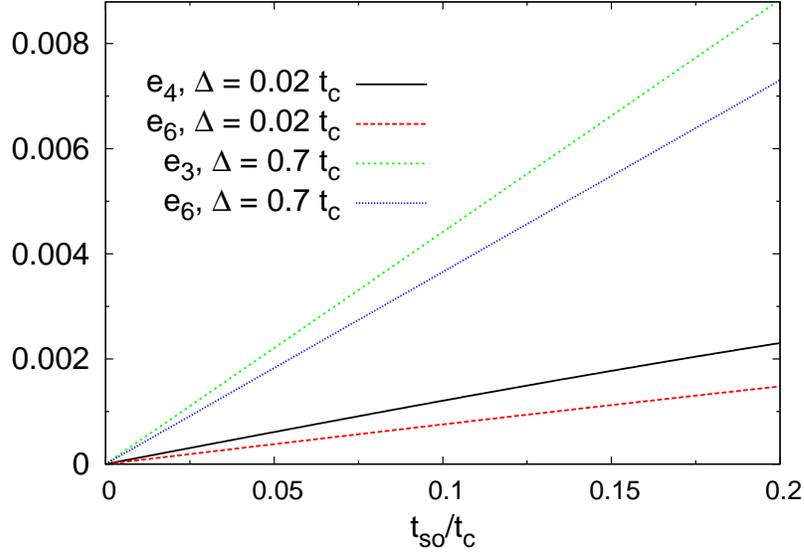}
\caption{ (Color online) The figure shows the absolute amplitudes
$e_{3}$, $e_{4}$ and $e_{6}$ which express the mixing of the
triplets $\ket{T_{+}}$ and $\ket{T_{-}}$ with the $\ket{S_{02}}$
state. The Zeeman splitting is chosen in the low field regime
$\Delta=0.02t_{\mathrm{c}}$ and asymptotic regime
$\Delta=0.7t_{\mathrm{c}}$.}\label{amplit}
\end{center}
\end{figure}

Equation~(\ref{I0}) can be used to estimate the width of the
current peak that occurs at $\Delta_{r}=g_{e}\mu_{B}B_{r}$.
Specifically, if we denote by $\Delta_{r}-\Delta_{1}$ the half
width at half maximum of the peak, then $\Delta_{1}$ satisfies the
relation $I_{0}(\Delta_{1})=I_{r}/2$. In the same way the width of
the $\Delta=0$ minimum can be estimated in the regime $I_{a}\sim
I_{r}$. If $\Delta_{1}$ corresponds to half of the width then to a
good approximation it has to satisfy the relation
$I_{0}(\Delta_{1})=I_{a}/2$.

The variation of the current as a function of the magnetic field
and SOI strength is mainly due to the change of the first terms in
$E_{i}$ and $F_{i}$. For example, if $t_{\mathrm{so}}\ll
t_{\mathrm{c}}$, then as can be seen in Fig.~\ref{parameters} only
the quantities $(\gamma_{1}d_{4})^2$, $(\alpha_{2}a_{6})^2$, and
$(\gamma_{1}d_{3})^2$ which contribute to $E_{i}$ and $F_{i}$ are
important. Because these quantities are approximately equal we
derive that $E_a \approx E_{0}$ and $F_a \approx F_{0}$, thus
$I_{a}\sim I_{0}$. In the same way, at $B_{r}$ the corresponding
amplitudes in Eqs.~(\ref{er}) and~(\ref{fr}) lead to $E_{0} >
E_{r}$ and $F_{0}
> F_{r}$. Thus, $I_{r}> I_{0}$ and a peak is formed at $B_{r}$.
On the other hand, in the range $t_{\mathrm{so}}< t_{\mathrm{c}}$
the quantities $(\delta_{1}e_{3})^2$, $( \gamma_{2}d_{3}+\beta_{2}
e_{3} )^2$, and $(\beta_{2}e_{6})^2$, $( \alpha_{1}a_{6} -
\delta_{1}e_{6} )^2$ increase drastically (Fig.~\ref{parameters}).
As a result, $E_{a}$ and $F_{a}$ decrease significantly, whereas
$E_{0}$ and $F_{0}$ do not change a lot. Therefore, for
intermediate or large $t_{\mathrm{so}}$ the asymptotic current
$I_{a}$ is much larger than the current at very low fields and it
approaches the current $I_{r}$. Eventually as $t_{\mathrm{so}}$
increases the current shows a dip for $\Delta=0$.

Inspection of the various amplitudes involved in the transition
rates demonstrates that the important amplitudes in order to
understand the current are the $e_{4}$ (or $e_{3}$ when $B$ is
high) and $e_{6}$. These amplitudes express the mixing of the
$\ket{T_+}$ and $\ket{T_-}$ states with the $\ket{S_{02}}$
component. These amplitudes are responsible for the behaviour of
the current when $t_{\mathrm{so}}$ is included in the model. As
shown in Fig.~\ref{amplit}, when $t_{\mathrm{so}}$ is not too
small the mixing of the $\ket{T_+}$, $\ket{T_-}$ states with the
$\ket{S_{02}}$ component is stronger in the asymptotic regime than
for $B\sim0$. For this reason the current is more sensitive to
$t_{\mathrm{so}}$ when $B$ is high but shows only a small
variation with $t_{\mathrm{so}}$ for $B\sim0$. Furthermore, as B
increases (in the asymptotic regime) then for a fixed
$t_{\mathrm{so}}$ and negative detuning the amplitude $e_{6}$ for
the $\ket{T_{+}}$ state increases, whereas the amplitude $e_{3}$
for the $\ket{T_{-}}$ state decreases. This behaviour can be
understood by noticing that the corresponding anticrossing points
move along the detuning axis in opposite directions. In
Eq.~(\ref{Ia}), $E_{a}+F_{a}$ is to a good approximation constant
and the current $I_{a}$ remains constant as $\Delta$ increases.

\begin{figure}
\begin{center}
\includegraphics[height=11.50cm,angle=270]{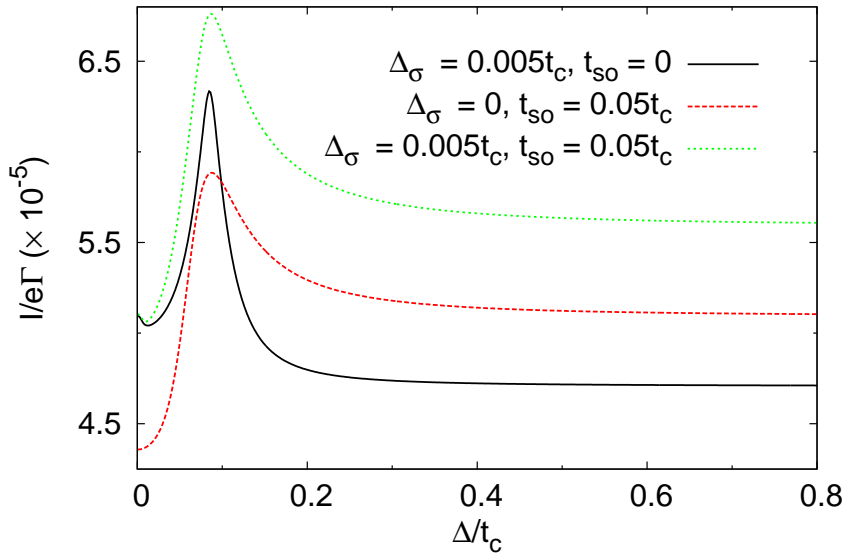}
\caption{ (Color online) Electrical current, $I$, as a function of
the Zeeman splitting, $\Delta$, for the energy detuning
$\delta=-25t_{\mathrm{c}}$ and different combinations of SOI and
HI strengths. Here $\Delta_{\sigma}=
g_{\mathrm{e}}\mu_{\mathrm{B}}\sigma_{\mathrm{N}}$.}\label{hicurr}
\end{center}
\end{figure}

To take into account the hyperfine interaction, we follow a
standard approach and treat the nuclear magnetic fields ${\bf
B}_{\mathrm{N},i}$ in the two dots as quasi-static classical
variables which take random values.~\cite{jouravlev} In this case
the electron and nuclear spin dynamics are uncoupled. In the next
section we develop a model to look at the coupled dynamics. The
distribution for each random static field is Gaussian with spread
$\sigma_{\mathrm{N}}$. The electrical current is computed as the
average over different random field
configurations.~\cite{jouravlev} This is a good approximation when
the nuclear dynamics is much slower than the electron dynamics. In
Fig.~\ref{hicurr} we plot the current versus Zeeman splitting.
When $t_{\mathrm{so}}=0$, a peak is formed at the singlet-triplet
anticrossing point due to the mixing caused now by the HI. When
both $t_{\mathrm{so}}$ and $\sigma_{\mathrm{N}}$ are nonzero the
resonant and the asymptotic current increase. However, for
$\Delta=0$ the current is determined by the HI, consistent with
the results in Ref.~\onlinecite{danon2009}. As the spread
$\sigma_{\mathrm{N}}$ increases, the HI results in a peak at
$\Delta=0$, though the $\Delta=0$ current may have a more
complicated form when $\sigma_{\mathrm{N}}$ is large. In the small
$\sigma_{\mathrm{N}}$ regime that we are interested in and for
$t_{\mathrm{so}}\neq 0$, the $\Delta=0$ current is typically
smaller than the asymptotic current. Our numerical calculations
confirm that these trends occur for different choices of
$t_{\mathrm{so}}$ provided that
$g_{\mathrm{e}}\mu_{\mathrm{B}}\sigma_{\mathrm{N}} \ll
t_{\mathrm{so}}$.

\section{Nuclear spin polarization and transient dynamics}

\subsection{Physical model}

The model employed in the previous section provides insight into
the current, but it does not capture the coupled electron-nuclear
spin dynamics. Thus, for example, the nuclear spin polarization as
a result of the transported electrons through the DQD cannot be
examined. In this section we look at the DQD system from a
different perspective. Specifically, we use an idealized model to
study the coupled electron-nuclear spin dynamics and how this
affects the transient behaviour of the current. We will use this
model to show two things. First, that the presence of even a weak
spin-orbit coupling can prevent nuclear spin polarization. Second,
if the nuclear spin state is completely thermalized several
interesting features arise in the transient electron current;
regular beating followed by quasi-chaotic oscillations.

A true model of the $2^N$ states which describe the $N$ nuclear
spins is computationally intractable, but some insight can be
gained from, e.g., treating the nuclear spins as a ``giant''
spin.~\cite{Rudner10, Brandes, Inoshita2002, Inoshita2004} In
addition to reduce the complexity even further, we restrict
ourselves to a small subspace of the two-electron Hilbert space,
spanned by the states $\ket{T_+}$, $\ket{S_{11}}$, and
$\ket{S_{02}}$. This is a reasonable approximation under an
appropriate bias, i.e., when only the state $\ket{T_+}$ is in the
voltage bias window (and neglecting tunneling, from the
reservoirs, into superpositions of the singlet states).  In
addition we assume that the rate of tunneling from the left lead
to the dot is large, and that we are at the anti-crossing point of
the singlet-triplet states.~\cite{danon2009} This reduction of the
state-space does also neglect the occupation of the
$\ket{0,\downarrow}$ state, and trapping in other
single-electron-occupation states. As a test, we included the
single-electron state, $\ket{0,\downarrow}$, in an extended model,
but it had little impact on the results we present here. Thus
hereafter it will be neglected. In addition, another interesting
regime to investigate would be to assume a larger bias and include
all three triplet states. However, since all the interesting
coupled electron-nuclear effects we discuss here are mediated by
the $\ket{T_+}$ and $\ket{S_{11}}$ dynamics, such a regime may
only reduce the visibility of these effects.

We model the interaction of these three states with a
non-equilibrium master equation. This model allows a flow of
electrons through the DQD, and thus we can estimate properties
like the total polarization of the nuclear spin, and the transient
dynamics of the coupled electron-nuclear spin system. This is in
contrast to treating the spins as a frozen spin bath,~\cite{Rosen}
as a semi-classical degree of freedom,~\cite{Brandes, Rudner2011}
or as a non-Markovian environment.~\cite{Ru}

The total Hamiltonian of the system is then written as \beq H =
H_{\mathrm{c}} + H_{\mathrm{so}} +H_{\mathrm{hf}}- \Delta
\ket{T_+}\bra{T_+} - \delta \ket{S_{02}}\bra{S_{02}},\eeq where
$H_{\mathrm{c}}$ describes the coupling between the singlet states
\beq H_{\mathrm{c}}=
t_{\mathrm{c}}\big[\ket{S_{11}}\bra{S_{02}}+\ket{S_{02}}\bra{S_{11}}\big],
\eeq and $H_{\mathrm{so}}$ is the Hamiltonian part due to the SOI,
which couples singlet-triplet states, and has the form \beq
H_{\mathrm{so}} =
t_{\mathrm{so}}\big[\ket{T_+}\bra{S_{02}}+\ket{S_{02}}\bra{T_+}\big].
\eeq The hyperfine interaction is \beq{H_{\mathrm{hf}}=
\frac{g}{2} \sum_{k=1}^N \left[\ket{T_+}\bra{T_+} I_z^k +
\frac{1}{\sqrt{2}}\left\{\sigma^{(1)}_- I_+^k+\sigma^{(1)}_+
I_-^k\right\} \right]}.\label{Hn} \eeq  Here $\delta$ is the
energy detuning and $\Delta=g_e\mu_{B}B$ is the Zeeman splitting
caused by the external magnetic field $B$ in the $z$-direction,
$t_{\mathrm{so}}$ is the spin-orbit coupling, and $t_{\mathrm{c}}$
is the coherent tunnelling amplitude for the singlets. In
$H_{\mathrm{hf}}$ we use an effective-spin notation so that
$\sigma^{(1)}_+ =\ket{T_+}\bra{S_{11}}$. The coupling $g$ is the
nuclear hyperfine coupling term, $g=A/N$, where $N$ is the number
of nuclear spins and $A$ is in the range of 80 $\mu$eV.

For simplicity, we assume the nuclear spins are spin-1/2, and the
hyperfine coupling strength is uniform and homogeneous. This
implies equal-size dots, and an equal nuclear hyperfine coupling
on each site. We are working in a rotating frame and the sign
difference between couplings on the left and right dots, due to
the antisymmetry of the wavefunction, is omitted.~\cite{Rudner10}
Thus the nuclear states are in fact the difference between nuclear
spin states in the left dot and the right dot. This allows us to
use the giant-spin approximation $ \hat{J}_i = \sum_k I_i^k, $
$i=z,+,-$. For a large thermal state one should really describe
the spin system as a distribution over giant spins of differing
sizes.~\cite{Rudner10} Here we consider a single giant spin of
length $J= \sqrt{N/2}$, which may be valid if the distribution of
spin sizes is strongly peaked (see Ref.~\onlinecite{Rudner10} for
a rigorous discussion of this assumption). In the final section we
discuss the possible effects of a broader distribution.

To account for electron transport  we include a Lindblad term,
\beq L_1[\rho]= \frac{\Gamma}{2} \left [ 2 \mu_+ \rho \mu_- -
\mu_- \mu_+ \rho - \rho \mu_- \mu_+ \right], \eeq where the
electron transport operator is $\mu_- =
\ket{S_{\mathrm{02}}}\bra{T_+}$, and we have omitted the vacuum
state which is a good approximation when the rate of tunnelling-in
from the left lead is large. Thus, in the results that we show in
the following figures, we solve the master equation for the
combined electron-nuclear spin density matrix $\rho$, \beq
\frac{d\rho}{dt} = -\frac{i}{\hbar} [H,\rho] + L_1[\rho]. \eeq
Because of the small Zeeman splitting of the nuclear spins, we
impose an effective infinite temperature for the nuclear system
initial state. Nuclear spin dephasing and thermalization are not
explicitly included here as they can, in principle, break the
large spin approximation, and cannot be written in terms of the
$J_z$ operators alone. Fortunately, the nuclear spins are
typically weakly interacting with each other and the environment.
We assume that the current is dominated by transport into the
right reservoir, and is defined as \beq I(t) = e \mathrm{Tr}[
\mathcal{J_R} \rho(t)],\eeq where the superoperator is the jump
operator $\mathcal{J_R}[\cdot]=\Gamma\mu_+ \cdot \mu_-$. For the
polarization of the nuclear spins we calculate the expectation
value of $J_z$ for the large spin.

\begin{figure}[]
\begin{center}
\includegraphics[height=11.5cm]{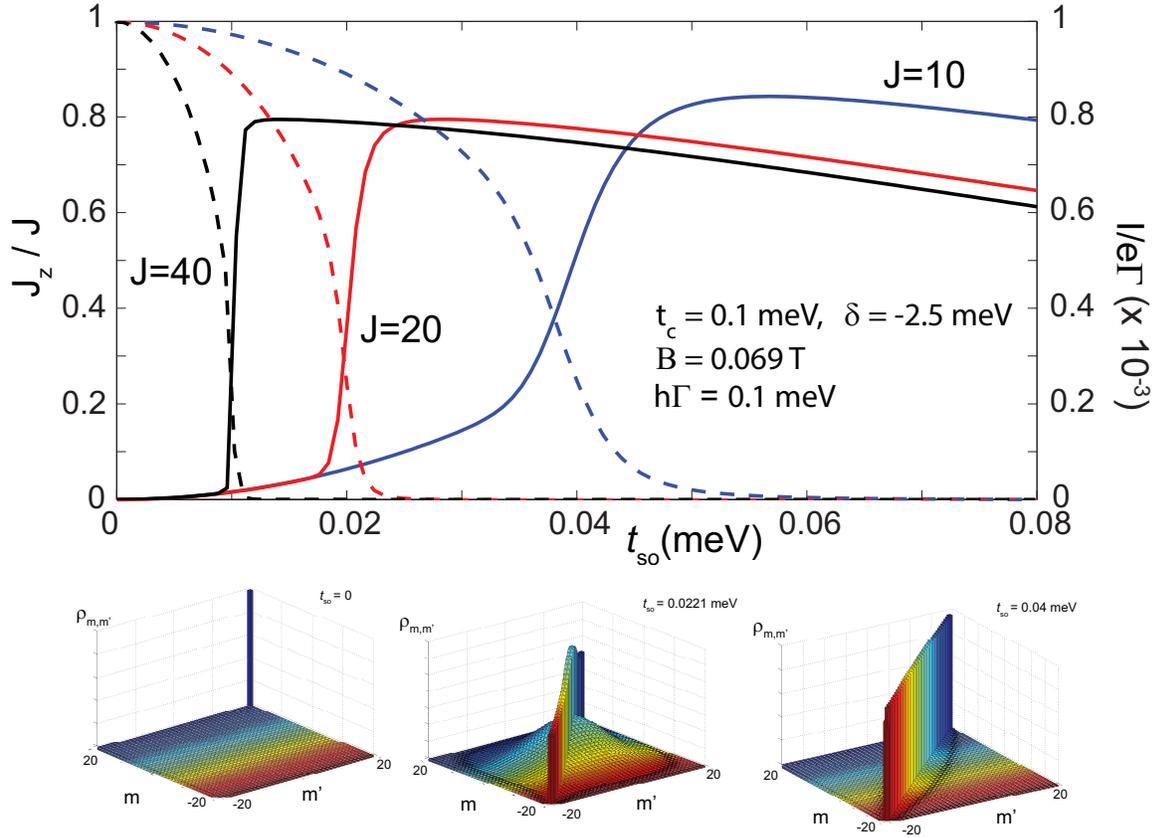}
\caption{\label{lev} (Color online) Top: Steady-state current,
$I$, (solid lines) and nuclear spin polarization, $J_z$, (dashed
lines) as a function of $t_{\mathrm{so}}$.  Three different sizes
of large spin are shown, $J=10$ (blue), $J=20$ (red), and $J=40$
(black), corresponding to $N=200$, $400$ and $3800$ nuclear spins
respectively. The transition occurs for $t_{\mathrm{so}} > 25
A/4\sqrt{N}$, where $A$ is the maximum value of the nuclear
hyperfine coupling $g=A/N$.  Because of the inverse scaling of the
transition point, as a function of $N$, these results suggest that
a relatively small spin-orbit coupling can suppress the
polarization of the nuclear spin. Bottom: matrix elements (in the
$m$ basis of the $J_z$ operator) of the steady-state reduced
density matrix of the nuclear spin, for $J=20$. At the critical
point the steady-state contains significant non-zero coherences.}
\end{center}
\end{figure}

\subsection{Nuclear spin polarization}

Hysteresis in the current measurement, as one sweeps the external
magnetic field through the singlet-triplet level crossing, is a
sign of nuclear spin polarization. In vertical dots, large
polarizations ($>40\% $) have been achieved.~\cite{Baugh,
Takahashi, Kobayashi, Petta} In lateral dots, the polarizations
are significantly smaller, perhaps due to either the asymmetry in
the dots, and thus in the nuclear hyperfine interaction, or dark
states.~\cite{Gullans} However, even for small levels of nuclear
spin polarization, large hysteresis has been observed. Further, in
several recent studies it has been shown theoretically that if the
spin-orbit coupling is above a certain threshold, relative to the
nuclear hyperfine coupling, no polarization of the nuclear spin
occurs.~\cite{Rudner10} If it is below that threshold, the nuclear
system becomes polarized due to the spin-flip process during the
electron transport. At some critical value between these two
regimes, long-lived dark states can occur, alongside a topological
phase transition.

To investigate this phenomenon in our model we plot, in
Fig.~\ref{lev}, both the steady-state current $I$ and the
normalized nuclear polarization $J_z/J$ as a function of
$t_{\mathrm{so}}$ for a given nuclear hyperfine coupling $A=0.1$
meV. We use dot parameters which put us at the singlet-triplet
resonance point, and also omit the first term (the Overhauser
term) in Eq.~(\ref{Hn}).  We see that for \beq t_{\mathrm{so}} <
25A/4\sqrt{N},\eeq the nuclear spin is strongly polarized by the
electron transport process, and the current flow is low.
Conversely, as \beq t_{\mathrm{so}} > 25A/4\sqrt{N},\eeq the
spin-orbit mediated transport route becomes dominant, and the
nuclear spin is no longer maximally polarized. The transition
between these two regimes is consistent with the sharp change
observed in the ``topological phase transition''~\cite{Rudner10}.
The factor of $25$ arises from the amplitudes $C_{0,2}$ and
$C_{1,1}$ of the combined effective singlet state used in that
treatment $\ket{S} = C_{0,2} \ket{S(0,2)} + C_{1,1}\ket{S(1,1)}$.
Inspection of the bare eigenstates of the electron spin
Hamiltonian shows that this factor is, in general,  \beq
|(C_{1,1}/C_{0,2})| = |(\delta -\sqrt{4t_c^2+\delta^2})/2t_c|.\eeq
Thus the quantum dot system parameters can in practice also  be
tuned to sweep the axis of Fig.~\ref{lev}. In the bottom half of
Fig.~\ref{lev} we show the absolute value of  the matrix elements
(in the basis of the $m \in \{-J, J\}$ eigenstates of $J_z$) of
the steady-state nuclear spin density matrix for three choices of
spin-orbit coupling: zero, at the ``critical point'' and far above
the critical point. One can easily see that at the critical point
there are significant non-zero coherences in the nuclear spin
state, which may be related to the presence of long-lived ``dark
states''.~\cite{Rudner10} Our results indicate that precursors to
this ``topological phase transition'' exist even in the presence
of a full transport cycle and noisy environmental effects, akin to
the persistence of quantum phase transitions in non-equilibrium
systems.~\cite{Dalla}

\begin{figure}[]
\begin{center}
\includegraphics[width=\columnwidth]{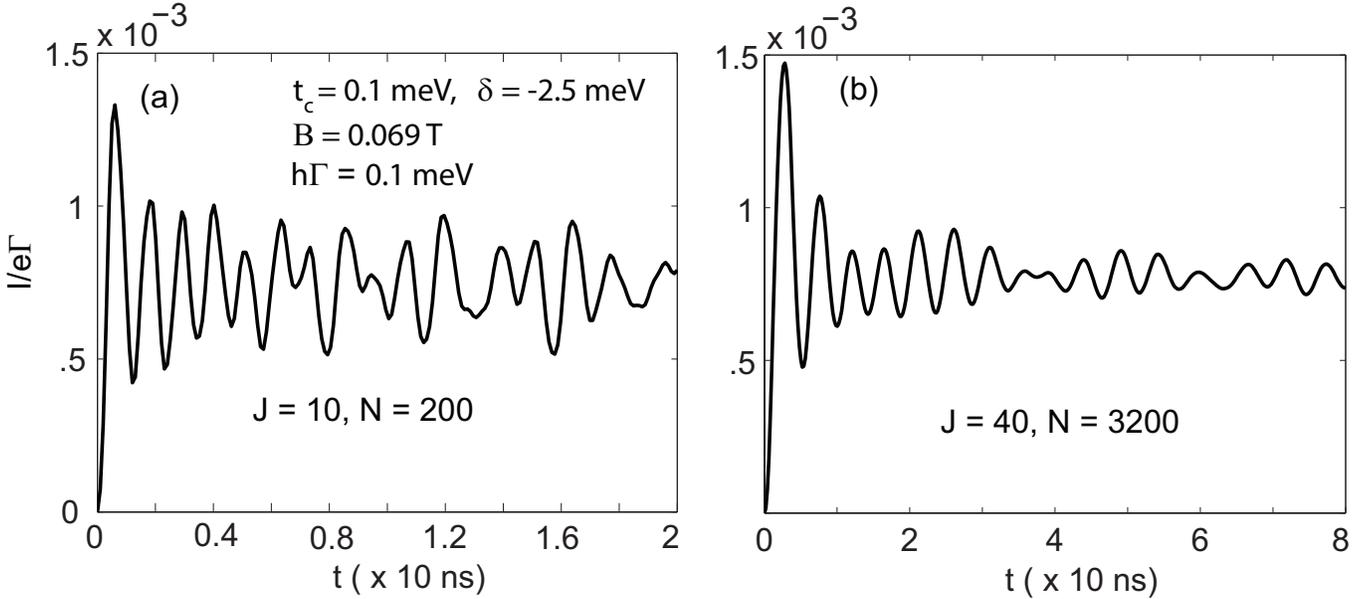}
\caption{\label{trans} Transient current, $I(t)$, with an
initially maximally-mixed nuclear spin state for (a) $J=10$, and
(b) $J=40$. The large number of commensurate frequencies around
$m=0$ cause the regular oscillations with frequency $A/4\sqrt{N}$,
before the onset of quasi-chaotic dynamics.}
\end{center}
\end{figure}

\subsection{Transient dynamics}

Some evidence~\cite{Takahashi, Reilly} indicates that the low
dephasing  and extremely long relaxation time of the nuclear
spins, combined with the fast stochastic electron transport
process inducing nuclear polarization, leads to a long time
instability phenomenon and fluctuations in the nuclear spin
state.~\cite{Brandes, Rudner2011} This is typically observed in
the long-time beating seen in the current.~\cite{Takahashi,
Reilly} This is almost certainly a semi-classical
effect,~\cite{ono2004, Brandes, Inoshita2004, Rudner2011, Danon}
though some evidence suggests that coherence within the nuclear
spins can survive for millisecond time scales.~\cite{Takahashi} To
gain some insight on what might be observed on shorter time
scales, we now examine the transient behaviour of the electron
transport induced by the nuclear spin. In the following we
entirely neglect the spin-orbit coupling.

We assume that at some initial time $t=0$ the electronic system is
prepared in the $\ket{T_+}$ state and the nuclear system is in an
initially maximally-mixed state \beq \rho(0) =
\frac{1}{2J+1}\sum_m \ket{m}\bra{m} \otimes
\ket{T_+}\bra{T_+}.\eeq We then solve the dynamics given by this
initial state in the equation of motion. As we will show below, we
find that this produces dynamics akin to the Jaynes-Cummings
Hamiltonian from quantum optics with an initial thermal (or
chaotic) cavity state.~\cite{Klimov, Knight}  This is a
well-studied system, with so-called ``quasi-chaotic revivals''. In
the optical case, studies have shown that one observes an initial
sharp peak~\cite{Knight} in the atomic state on a time-scale
$t_p^{(\mathrm{bosonic})} \approx
2\pi\hbar/\lambda\sqrt{\bar{n}}$, where $\lambda$ is the
field-atom coupling in that model, and $\bar{n}$ is the initial
thermal occupation of the field given by the Bose-Einstein
distribution. This is typically followed by the onset of the
so-called ``quasi-chaotic'' oscillations at a time that scales
with the thermal occupation.

In Fig.~\ref{trans} we show the transient electron current for two
sizes of the nuclear spin system. Again we omit the Overhauser
term.  In the figure, one can see the clear onset of quasi-chaotic
beating and a scaling of the onset of this beating with the size
of the large spin $J$. Since we are now dealing with an
infinite-temperature large spin, the role played by $\bar{n}$ in
the optical case is now played by $J$ in the large-spin case.
Unlike in the optical case, however, the collapse of an initial
peak is not seen here. Instead, there is a transition from large
steady oscillations to quasi-chaotic ones.

To understand the presence of the large steady oscillations  we
can make two observations. Firstly, for low thermal occupation
$\bar{n} \ll J$, and large $J$, the Holstein-Primakoff
transformation tells us that the dynamics would coincide with that
of the bosonic Jaynes-Cummings model, but with renormalized
coupling $\lambda \approx A/2\sqrt{2}N$.  As $\bar{n} \propto J$,
the analogy should break-down. Secondly, the large regular
oscillations can be understood by explicit diagonalization of the
$H_{\mathrm{hf}}$ Hamiltonian (again omitting the first,
Overhauser, term), which gives the following expression for the
occupation probability of the $\ket{T_+}$ state, assuming that the
initial state is $\ket{\psi(0)} = \ket{T_+} \otimes
\sum_{m=-J}^{J} C_m \ket{m}$, \beq |\langle T_+|\psi(0)\rangle |^2
= \sum_{m=-J}^{J} |C_m|^2 \cos^2\left(\frac{E_+
t}{\hbar}\right),\eeq where \beq E_+^{(m)} = \frac{A \sqrt{J(J+1)
- m(m+1)}}{2\sqrt{2}N}.\label{e+}\eeq In the bosonic case, this is
an infinite sum whose frequencies scale as $\sqrt{n+1}$. One can
understand the large regular oscillations that occur in the large
spin case by considering the contributions to the sum around
$m=0$. This region contributes a large number of terms to the sum,
with similar frequencies $E_+^{(m\approx 0)}/\hbar \approx
A/4\hbar\sqrt{N}$, which are commensurate at small times. This
gives rise to the frequency of the early-time oscillations in
Fig.~\ref{trans}.

The possibility to observe both the early-time oscillations and
the quasi-chaotic oscillations in experiment is intriguing. One
can consider that as $N$ is increased the period of the
oscillations increases. For $N\propto  10^5$--$10^6$ this can
reach the order of 100 micro-seconds. Ultimately, the observation
of these oscillations is limited by two effects. First, electron
spin dephasing will affect these dynamics. We found that, by
including such effects in our model, if these dephasing rates are
much larger than the frequency $E_+$ the oscillations become
damped. However, the oscillations are not strongly affected by
dephasing or decoherence in the charge degree of freedom.
Secondly, in reality, the width of the distribution of large spins
in the thermal ensemble around $\sqrt{N/2}$ will also cause
dephasing, and is the primary cause of the electron spin dephasing
to begin with. One can consider that large spins in this
distribution close to $\sqrt{N/2}$ will contribute ``in phase'' to
the early-time oscillations and those far away will induce
additional dephasing. Thus, inevitably the oscillations shown in
Fig.~\ref{trans} will decay if this distribution is broad.
Finally, dephasing and rethermalization of the nuclear spin states
will also affect the dynamics if these rates become comparable to
$E_+$.

In addition, the behavior we show in the dynamics of the current
occurs on a relatively short time scale and strongly depends on
the initial state. We also emphasize that the meaning of this
short-time dynamical current we plot in the figures is the
following: it is the ensemble average of detecting the current of
a single electron entering the reservoir based on a specific
initial condition. In other words, in a real experiment, one would
have to repeatedly prepare the nuclear-spin and dot system in the
same initial state, and measure the resulting single electron
transport events to eventually collate the data shown in the
figure. Obviously this is experimentally challenging.  A more
feasible and natural approach would be to detect transient
behavior in the high-frequency current-noise spectrum. However,
given that such transients are ``around the steady state'' (a
state which may include significant nuclear spin polarization),
some of the features which rely on the nuclear spin state being in
a maximally-mixed state may become less visible.

In practice, it maybe more feasible to consider a closed system,
i.e., disconnected from electronic reservoirs, and measured by,
e.g., a charge detector. In this case there will be no dynamical
nuclear polarization, which is advantageous for observing the
features which rely on the maximally mixed state.  Also, at this
stage, it is not clear if there is any connection between the
oscillations we observe here and those seen in experiments.  Even
for large $N$ the time scales still differ greatly.  Finally, an
alternative system to investigate this phenomenon could be quantum
dots in carbon nanotubes, or with superconducting qubits/wave
guides~\cite{nature} coupled to ensembles of spins in diamond.

\section{Conclusions}

This work investigated electronic transport in a double quantum
dot for weak spin-orbit interaction, namely, when the spin-orbit
amplitude is smaller than the interdot coupling. The electrical
current was calculated numerically from rate equations. We derived
simple approximate expressions for the current as a function of
the amplitudes of the one- and two-electron states which
participate in the transport cycle. We found that when the SOI is
small the current shows a peak at a magnetic field for which the
lowest two-electron energy levels anticross, whereas when the SOI
is large a dip is formed at zero magnetic field. Numerical
calculations showed that in a double dot system with a small
hyperfine interaction these characteristics remain valid.

We also considered a model which includes dynamical behavior of
the nuclear spin, and investigated the coupled dynamics between
electron and nuclear spins. We found that the conflict between the
spin-orbit and nuclear hyperfine couplings results in a transition
point between no polarization and large polarization of the
nuclear spin ensemble. We also considered the transient dynamics,
where the nuclear spin ensemble is initially prepared in a highly
thermal state. The unique characteristics of the large
effective-spin model used to describe the nuclear spin ensemble
induces dynamics in the current which depends on the fundamental
nuclear hyperfine coupling and the ensemble size.

\section*{ACKNOWLEDGEMENTS}

We would like to thank K. Ono, S. Amaha, J. R. Johansson and
especially M. Rudner for useful discussions. G.G. acknowledges
support from the Japan Society for the Promotion of Science (JSPS)
No.~P10502. F.N. is partially supported by the ARO, NSF grant
No.~0726909, JSPS-RFBR contract No.~12-02-92100, Grant-in-Aid for
Scientific Research (S), MEXT Kakenhi on Quantum Cybernetics, and
the JSPS via its FIRST Program.


\begin{thebibliography}{5}
\bibitem{newreference}
Numerical results derived from the time evolution of the full
density matrix confirm that this is a very good approximation,
even when degenerate eigenvalues occur, provided the coupling to
the leads is weak. We also ignore cotunneling processes.
\end{thebibliography}
\end{document}